\documentclass[aps,prd,amsmath,floats,floatfix,superscriptaddress,nofootinbib,showpacs,reprint]{revtex4}

\usepackage{hyperref}

\usepackage{graphicx}
\usepackage{xspace}
\usepackage[usenames,dvipsnames]{color}
\usepackage{amsmath,amssymb}
\usepackage{longtable}
\usepackage{ulem} 

\newcommand{\UNC}{\affiliation{Department of Physics and Astronomy,
    UNC-Chapel Hill, NC 27599, USA}}

\begin{document}

\title{Predictions of the Quantum Landscape Multiverse}

\author{Laura Mersini-Houghton} \UNC 
\date{\today}

\begin{abstract}

 The 2015 Planck data release has placed tight constraints on the class of inflationary models allowed. The current best fit region favors concave downwards inflationary potentials, since they produce a suppressed tensor to scalar index ratio $r$. Concave downward potentials have a negative curvature $V''$, therefore a tachyonic mass square that drives fluctuations. Furthermore, their use can become problematic if the field rolls in a part of the potential away from the extrema, since the semiclassical approximation of quantum cosmology, used for deriving the most probable wavefunction of the universe from the landscape and for addressing the quantum to classical transition, breaks down away from the steepest descent region. We here propose a way of dealing with such potentials by inverting the metric signature and solving for the wavefunction of the universe in the Euclidean sector. This method allows us to extend our theory of the origin of the universe from a quantum multiverse, to a more general class of concave inflationary potentials where a straightforward application of the semiclassical approximation fails. The work here completes the derivation of modifications to the Newtonian potential and to the inflationary potential, which originate from the quantum entanglement of our universe with all others in the quantum landscape multiverse, leading to predictions of observational signatures for both types of inflationary models, concave and convex potentials.  

\end{abstract}


\maketitle

\section{Introduction}

Once again Planck has spoken. The 2015 release \cite{planck1,planck2}, which includes polarization data, tightened the constraints on the plethora of the allowed inflationary models. At the same time it strengthened the evidence for the existence of some of the CMB anomalies. The best fit region for inflationary models is now that of concave downwards potentials, such as the Starobinsky $R^2$-model \cite {staro} and the hilltop class of models \cite{hilltop}. The reason for favoring concave downward potentials is simple: they produce a suppressed tensor to scalar ratio $r$. The tensor to scalar ratio is tightly constrained by Planck 2015 \cite{planck1} to be $r \le 0.07$. Since $r$ is related to the amplitude of scalar fluctuations $A_s$, the new constraints in $r$ have induced a small change in $A_s$ relative to its 2013 reported value \cite{planck2013}. Consequently, bounds on $A_s$ constrain the energy scale of inflation at the start of slow roll $V_0$, and affect the significance of the low multipole power suppression in temperature autocorrelations. However the accuracy of measuring the power at low multipoles is partially impeded by cosmic variance. The 2015 data \cite{planck2, planck2013} also provides additional support
 and information about the anomalies, increasing their significance to $4 \sigma$ in some cases, such as the cold spot.

 The big mystery from Planck 2013 data, namely the intriguing friction between the best fit inflationary models and the anomalies' best fit region, persists in the Planck 2015 data \cite{planck2}. The parameters favored by the anomalies are $4 \sigma$ away from the best fit region of inflationary models, and vice versa. The new hints and constraints from the Planck satellite data \cite{planck1,planck2} help us to further probe the physics of the universe as it came into existence and, possibly, even before it emerged. Understandably, the search for a coherent theory with predictive powers, that simultaneously accommodate the constraints placed on the inflationary models and on the anomalies, is well motivated.

\section{The Quantum Landscape Multiverse}
\label{sec:model}

In 2004, when the string theory landscape \cite{landscape} was announced, I proposed that we take the landscape to be the space of the initial states of the universe, and allow the wavefunction of the universe to propagate on this landscape \cite{archillmh} in order to address the selection of the initial conditions for our universe through a concrete calculation by means of quantum cosmology. The purpose of the proposal was to find a way to derive, from first principles rather than by a postulate, the selection criterion for the initial state of the universe based on the dynamics of the system on a superspace of initial states. The three assumptions made were: firstly, the formalism of quantum theory, including quantum cosmology, can be relied upon in these regimes; secondly, the vast landscape provides the space of the initial conditions; and, thirdly that the semiclassical approximation is valid. In this theory, the most probable universe is derived from the solutions to a {\it generalized} Wheeler DeWitt equation \cite{richlmh} for the wavefunctional of the universe propagating through landscape energy 'valleys'.

The solutions for the wavefunction of the universe derived in this approach, taught us that: the landscape can not be reduced to a quantum double well otherwise important quantum effects which induce localization of the wavefuntion and quantum interference, are not captured. The landscape has to be treated as a quantum $N-$ body problem; and the fine details of the distribution of vacua on the landscape do not matter as long as this distribution is disordered. Solutions for the wavefunction of the universe depend only on the strength of the disorder, the dimensionality of the landscape, and on the boundary conditions. So, in a sense, these solutions are universal, since any disordered quantum gravity 'landscape' of the same dimensionality, including that from string theory, would yield the same family of solutions independently of the detailed distribution of their respective vacua, for as long as they had the same disorder strength. On the other hand, any perfectly ordered landscape, such as the SUSY sector of the landscape \cite{archillmh}, cannot produce classical universes, since the wavefunction of the universe solutions for periodic potentials are not localized around a single vacua - rather they are of the 'Bloch waves' type, extended over the whole landscape. Therefore, a disordered landscape seems to be a generic requirement of any theory of quantum gravity that aims to explain the emergence of a quantum to a classical universe. The requirement of a 'disordered landscape' from quantum gravity implies that only two of the three assumptions in the list above are needed for deriving the selection of the initial conditions of our universe and testable predictions for the theory. The two assumptions are: the validity of quantum cosmology, and the validity of the semiclassical approximation.

Decoherence was included in this proposal in 2005 \cite{richlmh}, by considering the backreaction of long wavelength fluctuations comprised of fluctuations around the landscape vacua and the metric of $3$ geometries.The long wavelength fluctuations make up the 'environment', while the wavefunction branches comprise the 'system'. Including the effect of the 'environment', which contains an infinite number of flucutuations $f_n$ coupled weakly to the 'system', triggers decoherence among the wavefunction branches localized on landscape vacua. Decoherence is responsible for the derived selection mechanism of the most probable initial state. The selection mechanism emerges from the quantum dynamics of gravitational and scalar degrees of freedom. Solutions to the generalized Wheeler DeWitt equation, (WdW), now residing on an infinite size midi-superspace that included $f_n$, showed that the most probable universes select the high energy vacua on the landscape (see \cite{richlmh} for the detailed derivation).

Coherence and decoherence are closely related. Therefore we can use entanglement among the wavefunction branches to our advantage to derive a series of testable predictions. Using the semiclassical approximation, we calculated in \cite{tomolmh} the entanglement strength from the backreaction term which had triggered decoherence among the branches/wavepackets of the wavefunctional of the universe. We found that entanglement modifies the CMB and gravitational potential in our universe by contributing a nonlocal and scale dependent, quantum correction term to the inflation potential and an additional source for the CMB perturbations, which we then evolved forward in time to the present day. We calculated a series of predictions in 2006 \cite{tomolmh} by estimating the effect of entanglement of our branch-universe wavepacket, with all other surviving universes,  including: the existence of a giant void/cold spot, a lack of power at the lowest wavenumber $k \simeq 2$ leading to the hemisphere power asymmetry, a suppressed $\sigma_8 \simeq 0.8$, and, a giant void of about $10$ degrees at redhsift $z \approx 1$ now known as the 'cold spot'. The derivation of these signatures starting from a theory of the origin of the universe from the landscape multiverse, was computationally intense. Therefore we used a simple inflationary model, the exponential type in \cite{tomolmh}, to illustrate the theory of the origin of the universe from a quantum landscape multiverse, and to derive the predictions listed in \cite{tomolmh}.

We now know that the exponential potential and most other convex type potentials with $V'' >0$, are ruled out by Planck data because they predict a higher tensor to scalar ratio $r$ than the one constrained by Planck \cite{planck1}. Since the data favors concave downwards potentials $V'' <0$ which predict suppressed tensor perturbations \cite{katiefred}, we here calculate the effect of entanglement and derive the corrections to the inflaton potential and the gravitational potential of our universe, for the case of concave inflationary potentials with $V'' <0$. These corrections provide the source of modification for the CMB and gravitational potential of our universe which lead to a series of predictions for convex potentials similar to the anomalies derived in \cite{tomolmh} for convex potentials. These modifications can be tested against current data and we carry this analysis in the companion paper \cite{eleonora}. The derivation of the modification to the gravitational potential produced by the quantum entanglement of our universe with all others in the quantum multiverse, is considerably subtler for the case of concave potentials. The subtlety is due to the fact that fluctuations are driven by $m^2 = V'' <0$, and the validity of the semiclassical approximation based on a saddle point expansion of the action for the wavefunctional of the universe, becomes questionable when the slow rolling field is away from its vacuum state, as is the case for some of the concave downwards potentials favored by data. If we 'blindly' applied the results of \cite{richlmh,tomolmh} to the case of these concave potentials, in the region away from the saddle point and with fluctuations driven by a tachyonic mass $m^2 <0$, then the 'decoherence factor' among the wavefunction branches would appear to grow instead of being suppressed, which is clearly wrong. The latter demonstrates that using the semiclassical method to estimate the decoherence factor in a regime where the method breaks down is an incorrect procedure.

But, the predicted contributions from the entanglement of our universe with all other wavefunction branches to the CMB and the gravitational potential of the universe, provide a powerful way of testing the quantum origin of the universe from a landscape multiverse. We perform the calculation in this paper. We calculate the entanglement of our branch with all others for the case of inflationary concave potentials with $V''<0$, and derive its effect on the observables of our universe in the present sky. The analysis of these predictions for concave and convex potentials, against the data from the Planck satellite, is shown in companion papers \cite{eleonora}. The current treatment thus completes the study of the effect that quantum entanglement from the landscape has in modifying the observables in our sky, for both concave and convex inflationary potentials.

\section{Entanglement and Decoherence in the Quantum Multiverse}

In the theory of the origins of the universe from the landscape multiverse given in \cite{archillmh,richlmh, tomolmh}, the wavefunctional of the universe $\Psi$ propagates through the landscape vacua. The landscape is captured by a collective variable \cite{landscape}, the moduli field $\phi$ with a potential $V(\phi)$, consisting of a large number of vacua with energies randomly distributed. Details can be found in \cite{archillmh} and \cite{richlmh, tomolmh}. The wavefunctional $\Psi[a,\phi]$ is defined on a minisuperspace parametrized by the scale factor $a$ of 3-geometries with an FRW line element, and the landscape moduli $\phi$. The wavefunctional satisfies a Wheeler DeWitt (WdW) equation

\begin{equation} 
\left[ H_{g} + H(\phi) \right] \Psi[a,\phi] = 0
\label{wdw}
\end{equation}

sometimes known as the constraint equation. The conjugate momentum $p_{a}, p_{\phi}$ and variables $(a, \phi)$ are promoted to quantum operators. $H_{g}$ is the gravitational hamiltonian derived from the Hilbert-Einstein action, and $H(\phi)$ is the hamiltonian of a scalar field with kinetic energy, and potential energy $V(\phi)$. In our case the landscape structure provides $H(\phi)$ with $\phi$ the moduli field, and the landscape vacua energies are captured by the potential $V(\phi)$. Explicitly, rewriting the scale factor as $a=e^{\alpha}$ \cite{richlmh}, leads to the following expression for the total hamiltonian $H = H_{g} + H(\phi)$

\begin{equation}
H = \frac{1}{2e^{3\alpha}} \left[ \frac{4 \pi}{3 M_{p}^{2}} \frac{\partial^2}{\partial \alpha^2} - \frac{\partial^2}{\partial \phi^2} + V(\alpha \phi) \right],
\label{totalH}
\end{equation}

where $H_g$ is identified with the first term in Eqn.\ref{totalH}, and $V(\alpha, \phi) = e^{6 \alpha} V (\phi) - e^{4 \alpha} \kappa$, with $\kappa =0,1$ for flat or closed universes and $V(\phi)$ the landscape potential.
Note that $V(\phi)$ is the potential of a very large number ${\cal N}$ of vacua, a lattice of vacua sites, all with different energies described by a parameter $b$ which we can think of as the local $SUSY$ breaking parameter for each vacuum site. This approach and the probability distribution of the solutions for the wavefuncion of the universe, found from Eq.\ref{wdw}, (given in \cite{archillmh}) have not yet accounted for decoherence among the wavefunction branches, and for the quantum to classical transition. 

\subsection{Expanding around the Minimum of a Convex Potentials $V'' >0$}

To address decoherence among the branches of the wavefunction and therefore derive the emergence of a classical universe, we included an 'environment' comprised of long wavelength massive fluctuations labelled by { $f_{n}$ }. These are fluctuations around a vacua field $\phi = \phi_{0} + \Sigma_{n} f_{n}(a) Q_{n}$ with $Q_{n}$ the scalar field harmonics in the unperturbed metric of the 3 sphere,  and also perturbations around the 3-geometry FRW metric $h_{ij} = a^{2}(\Omega_{ij} + \epsilon_{ij}$ where $\Omega_{ij}$ is the FRW spatial metric and $\epsilon_{ij}$ is the perturbation around it (both scalar and tensor perturbations). The index $n$ is an integer denoting the mode number with physical wavenumber $k =n/a$.  The detailed procedure was laid out in \cite{richlmh}. The semiclassical approximation in \cite{tomolmh} was valid since the saddle point expansion in \cite{tomolmh} was performed near the minimum of a convex potential with $V''>0$. The long wavelength modes comprising the environment are the ones with wavelengths longer than the horizon. Thus superhorizon wavelength fluctuations are weakly (gravitationally) coupled to the 'system' or the branches of the wavefunction, localized on some vacua. The fluctuation modes are also independent of each other. Normally, both tensor and scalar fluctuations contribute to the 'environment' and their derivation is identical. But, the CMB fluctuation strength is of order $10^{-5}$ and tensor perturbations are much weaker than the scalar ones. So, the tensor fluctuations are orders of magnitude smaller than the scalar fluctuations, which is why we can focus on the effect of scalar fluctuations \cite{richlmh} only, without loss of generality \cite{kiefer,haliwell}.

The approach is based on the validity of the semiclassical approximation.The total action is expanded to second order around the saddle point, $S\simeq S_{0} + \frac{1}{2} \Sigma_{n} S"_{n} f_{n}^2$. In the WKB approximation, the total wavefunction can be written as $\Psi \simeq e^{i S/{\hbar}}$. Expanding the action around some vacua $\phi_j$, where for example the branch that becomes our universe is localized, we have $\phi^{j} \approx \phi^{j}_{0} + \phi^{j}_{n}(x,t)$ with $\phi^{j}_n = \Sigma f_{nlm}(a)Q^{nlm}(x)$. Here $(nlm)$ is collectively denoted by $n$, $Q^n$ are the scalar harmonics on a 3-sphere. We also drop the $j$- index from now by taking $\phi$ to be a continuous variable, since the number of vacua in the landscape is so large, about $10^{600}$ ,that we can safely consider $\phi$ to be a continuous instead of a discreet variable, in a nearly infinitely long lattice with a potential given by $V(\phi)$.

The total wavefunction can then be written as 

\begin{equation}
\Psi[a,\phi,f_{n}] = \Psi_{0}[a,\phi] \Pi_{n} \psi_{n}[a,\phi,f_{n}]
\label{fullwave}
\end{equation}

where $\Psi_{0} \simeq e^{i S_{0} /{\hbar}}$ is the unperturbed part of the wavefunction, $S_{0}$ the zero order term of the action evaluated at $\phi^{0}$, and $\psi_{n}$ denotes the contribution to the wavefunction from the perturbation modes. Of course we have to solve for $\psi_n$ and $\psi_0$.
Including these fluctuation modes results in an infinite sized midi-superspace parametrized by the $(a,\phi,f_{n})$. Their contribution generalizes the WdW equation \ref{wdw} into a 'Master Equation'. Using the expression \ref{fullwave} for the total wavefunction $\Psi$, inserting the perturbed metric $h_{ij}$ and field $\phi =\phi_{0} + \Sigma f_{n} (a) Q_n$, into the action, and expanding the action up to quadratic order around the saddle point, gives the hamiltonian modes $H_n$ for the contribution from $f_n$ perturbations to the total hamiltonian $H = H_{0} + \Sigma_{n} H_n$. At the quadratic level of expansion, these $H_n$'s are decoupled from each other \cite{kiefer,halliwell}.

The total wavefunctional of the universe then in the WKB approximation, up to the quadratic order expansion of the action, satisfies

\begin{equation} 
\left[ H_{g} + H(\phi) \right] \Psi[a,\phi] = \Sigma_{n} H_{n}(a,\phi,f_{n}) \Psi[a,\phi, f_{n}]
\label{master}
\end{equation}

where $H_n$ is the contribution to the hamiltonian from the individual modes $f_n$, i.e. the backreaction term added to the total hamiltonian of the WdW equation, obtained from the steepest descent expansion of the action. 

A semiclassical time parameter is defined from the action $\nabla S_{0} \nabla = \frac{\partial}{\partial t}$ where $\nabla$ is a derivative with respect to the minisuperspace variable $a$. The semiclassical definition of time is such that Einstein equations are recovered when the quantum to classical transition occurs, and the universe is a classical universe obeying general relativity equations, as shown in \cite{kiefer}. With this definition of the time parameter, plugging in the ansatz for the wavefunction Eq.\ref{fullwave} to the Master Equation Eq.\ref{master}, and the solution for the unperturbed wavefunction $\Psi_0$, yields an equation for the perturbations $\psi_n$, see \cite{richlmh}. These fluctuations obey a 'Schrodinger' type equation

\begin{equation}
H_{n} \psi_{n} = \frac{1}{a^3}\left[ -\frac{1}{2} \frac{\partial^{2}}{\partial f_{n}^{2}} + f_{n}^{2} \left(\frac{(n^{2}-1) a^4}{2} + \frac{m^{2} a^{6}}{2}\right)\right] \psi_{n} = i \frac{\partial \psi_{n}}{\partial t}
\label{fluctuationeq}
\end{equation}

Previously we expanded around the vacua of a convex potential \cite{richlmh} where the semiclassical approximation is valid, when calculating the backreaction term and predicting the series of signatures that test this theory. In \cite{richlmh} we considered the full wavefunction to be of the WKB form of Eq. \ref{fullwave}, with the unperturbed part of the wavefunction $\Psi_{0} = A_{0} e^{i \frac{S_{0}}{{\hbar}} }$ and took the ansatz for the perturbations to be

\begin{equation}
\psi_{n} = N(t)e^{i \frac{\Omega_{n}}{2} f_{n}^{2}}
\label{perturbation}
\end{equation}

The semiclassical approach leads to the emergence of a classical universe. The time parameter in Eq. \ref{perturbation} identified with $\nabla S_{0} \nabla = \frac{\partial}{\partial t}$ is the same parameter for all the branches.

With the inclusion of perturbations $\psi_n$, the WdW equation becomes the Master equation of Eq.\ref{master} and the sum over the perturbation hamiltonians $H_{n}$, given in Eq.\ref{fluctuationeq}, provides the backreaction term. From here on we will drop the index $n$ in $\Omega_n$ as a shorthand notation, but the reader should nto be confused into thinking all $\Omega_n$'s are the same, since they depend on $a$ and $n$. Inserting the ansatz for $\psi_{n}$ of Eq,\ref{perturbation} into the Schrodinger type equation for $\psi_n$ in Eq.\ref{fluctuationeq},gives the following two equations for $N(t)$ and $\Omega_n$ 
\begin{equation}
i \frac{dLnN}{dt} = Tr \Omega
\label{N}
\end{equation}
and,
\begin{equation}
-i \frac{\partial \Omega}{\partial t} = - \Omega^{2} + \omega^{2}
\label{omega}
\end{equation}

where $\omega^{2} = (n^{2} -1) + m^{2} a^{2} \approx n^{2} + m^{2} a^{2}$ for $n$ large, is the 'mass term' in the perturbation hamiltonian $H_n$ with $m^2=V''>0$. The notation $\Omega_{R} , \Omega_{I}$ denotes the real and imaginary parts of $\Omega$. The first, Eq.\ref{N}, can be easily integrated and we have
\begin{equation}
N(t) = det^{1/2}\left( \frac{2 \Omega_{R}}{\pi} \right) e^{-i Tr(\Omega_{R})}
\end{equation}
To solve Eq.\ref{omega} following \cite{kiefer} we used the ansatz
\begin{equation}
\Omega = -i a^3 \frac{{\dot y}}{y}
\label{omegaeq}
\end{equation}

which led to an equation for $y$
\begin{equation}
{\ddot y} +3 \frac{{\dot a}}{a} {\dot y} + (\frac{n^{2}}{a^2} + m^2) y =0
\label{lorentzy}
\end{equation}
that is solved by Bessel functions.

The reduced density matrix is obtained by tracing out the 'environment' degrees of freedom, i.e by integrating out the fluctuations $f_n$, as follows

\begin{equation}
\rho(a_{i} \phi_{i};a_{j},\phi_{j}) = \rho_{0} \Pi_{n} \int df_{n} \psi_{n}^{\star} (a_{j} \phi_{j} f_{n})\psi_{n}(a_{i} \phi_{i} f_{n})
\label{densitymatrix}
\end{equation}

where $\rho_{0} = \Psi_{0}(a_{i} \phi_{i}) \Psi^{\star}_{0}(a_{j} \phi_{j})$ is the unperturbed part. Once we have the solutions for $\psi_{n}$, we can calculate the reduced density matrix from Eq.\ref{densitymatrix}, which yields the following: $\rho = \rho_{0} det(\frac{\Omega_{R}}{\Omega})$.

From the reduced density matrix we obtain two crucial pieces of information: $i)$ how fast, the branches with 3-geometries $a_{i}, a_{j}$ localized on the landscape vacua $\phi_{i}, \phi_{j}$, are decohering from each other. (This information is given by the decoherence factor $Exp[-D]$, which is the real part in the exponent of the reduced density matrix Eq. \ref{densitymatrix} which shows how fast the cross term of the branches is suppressed, see below); $ii)$ and, how the backreaction from the 'environment' shifts the classical trajectories of the branches peaked around some energies $E_i$ in the midi-superspace, to a new value $E_i - \delta E_i$. As the branch undergoes a quantum to classical transition to become a universe, the energy shift around which a classical path in phase space is peaked, becomes a nonlocal correction to the inflationary energy and the gravitational potential of the universe, as shown in \cite{richlmh, tomolmh}.

Information about the energy shift $\delta E_{\phi}$ of the classical path of the branch, which is induced by interaction with the 'environment'  of $\Sigma_{n} H_n$, is contained in the imaginary terms of the exponent of the reduced density matrix Eq.\ref{densitymatrix}. It can also be calculated directly from evaluating $H_n$ as
\begin{equation}
\frac{<\Psi|\Sigma H_n \Psi>}{a^3} = \delta E_{\phi} = - \frac{1}{a^3}\left( Tr(\Omega_{R}) + Tr(\frac{{\dot \Omega_{I}}}{2\Omega_{R}}) \right)
 \label{energycorrection}
\end{equation}
 After the emergence of a classical universe,the shift in the classical path of the branch, which would have been peaked around the energy $V(\phi)$ if the backreaction term were ignored, corresponds to a shift of the inflaton energy $V(\phi) \rightarrow V_{eff}(\phi, b) = V(\phi) - \delta E_{\phi}$ that enters the Friedmann equation
\begin{equation}
3 M^2 {\cal H}^2 =  V_{eff}(\phi, b)= V(\phi) - \delta E_{\phi}
\end{equation}

where $M$ is the Planck mass and ${\cal H}$ the Hubble parameter.

Using the expansion
\begin{equation}
det (\frac{\Omega_{R}}{\Omega}) =e^{- Tr Ln[1+ i \frac{\Omega_{I}}{\Omega_{R}}]} \simeq e^{-i Tr(\frac{\Omega_{I}}{\Omega_{R}}) - \frac{1}{2} Tr (\frac{\Omega_{I}}{\Omega_{R}})^2 -...} 
\label{densityexpand}
\end{equation}

in the reduced density matrix allows us to immediately identify: the imaginary terms with the backreaction to the hamiltonian Eq.\ref{energycorrection}, $\frac{\Omega_{I}}{\Omega_{R}}$ with the interference length,  and the real term with the suppression factor among the branches, i.e. the decoherence factor $D = \frac{1}{2}Tr (\frac{\Omega_{I}}{\Omega_R})^2$. For the problem at hand, decoherence is dominated by $D\simeq \frac{\pi a m^3 H_{0}^{2}}{4 b^{2} }(a_i - a_j)^2$. We now need to apply these key steps in the derivation of the wavefunction of the universe and of quantum entanglement, to the subtlier case of concave downwards potentials, when the wavefunction is in a region where the semiclassical approximation breaks down.

\subsection{Concave Downwards Potentials, $V'' <0$}

All semiclassical approximations are based on saddle point evaluations of the WKB wavefunction $\Psi \simeq A e^{iS/{\hbar}}$, where $A$ is the amplitude and $S$ the action. Since $\Psi$ is oscillatory and $1/{\hbar}$ is large, then $\Psi$ and $\rho$ are dominated by points where $S' =0$ at $S=S_0$, and averages to zero otherwise due to the rapid phase oscillation. Therefore in the semiclassical approach we can write $\Psi \simeq A e^{\frac{i}{\hbar} (S_{0} + \frac{1}{2} S''\delta \phi^{2} +...)}$ where prime is a derivative with respect to the variable of $S$ around which it is expanded and $\delta \phi = f_{n}$. Problems with the semiclassical formalism arise if we move away from the saddle point region of the potential. For this regime we cannot justify the WKB ansatz for $\Psi$ of Eq. \ref{fullwave}. Also a 'semiclassical' time parameter needed to solve Eq. \ref{fluctuationeq} can not be defined. That is to say that the correspondence between the quantum wavefunction branches and classical universes is not a one to one mapping, there exist wavefunction solutions which have no classical universes counterpart.

We now know from the Planck results \cite{planck1} that the best fit to the data inflationary models are the concave downwards potentials  with $V''<0$, such as the Starobinsky type $R^2$- model \cite{staro} or the hilltop model \cite{hilltop}. In these potentials the field is slow rolling in a part of the potential which is far from the minimum. A blind application of the above semiclassical approximation is unjustified and leads to erronous results. If we were to apply this method naively it would lead to unstable fluctuations $f_n$ driven by $m^{2} = V''<0$ and an oscillatory or growing, instead of a suppressed decoherence factor, $e^{-D} \simeq e^{+ im \frac{m^2 H_{o}^{2}}{4 b^{2}}a (a_{i}-a_{j})^2 }$.

For these reasons, the case of concave potentials with $V''<0$, in region away from the steepest descent, is subtle and it is not clear how one can apply the semiclassical approach to these potentials. In terms of the landscape potential where the wavefunction propagates, this situation is similar to performing an expansion of the wavefunction when its branches localize somewhere near the plateau of the landscape potential barriers instead of being localized around its vacua.

In a different context, the authors of \cite{hawking} studied a similar case for AdS solutions of the wavefunctions of the universe, i.e. when $V<0$ and fluctuations are unstable $V'' <0$. Our case here is similar to theirs where fluctuations are concerned since we have $V''<0$,  but differs in the sense of the $3-$geometries being nearly DS spaces since we have $V>0$, instead of AdS geometries with $V<0$ studied in \cite{hawking}. This situation was further considered in \cite{vilenkin} who pointed out an important symmetry in quantum cosmology, namely: changing the signature of the metric $g_{ij} -> - g_{ij}$ is equivalent to inverting the potential $V(\phi) -> - V(\phi)$ thereby rotating to the Euclidean sector, and it is a symmetry of the WdW equation. Thus by inverting the potential we are rotating to the Euclidean sector where the fluctuations are stable since $-V'' >0$. We use this symmetry to our advantage and perform the whole semiclassical calculation for the perturbations and the decoherence factor in the Euclidean sector where the concave potential is inverted to to a potential well $V(\phi) -> - V(\phi) <0$. Fluctuations in the inverted potential are stable and driven by $m^2 = |V''| > 0$. At the end of the calculation we go back to the Lorentzian sector and translate our results in real time.

Let us now use the symmetry \cite{vilenkin} of the WdW equation $g_{ij} -> - g_{ij} , V -> -V$ to perform the perturbation expansion around the minimum of the inverted concave potential. In this case, the unperturbed wavefunction $\Psi_{0}= A_{0}e^{i S_{0}/{\hbar}}$ of the Lorentzian sector, goes to $\Psi_{0} = A e^{- \frac{S_{0}^{E}}{{\hbar}}}$ in the Euclidean plane. The Euclidean version of the 'Schrodinger' equation for the perturbations $\psi_{n}$ becomes

\begin{equation}
 \frac{\partial \psi_{n}}{\partial \tau} = \frac{1}{a^3}\left[ \frac{1}{2} \frac{\partial^2}{\partial f_{n}^{2}} - f_{n}^2 \left(\frac{(n^{2}-1) a^4}{2} + \frac{m^{2} a^{6}}{2}\right)\right] \psi_{n} 
\label{euclidfluctuationeq}
\end{equation}
 where we have the 'Euclidean' time parameter $\tau$ defined, by the Euclidean action $- S_{0}^{E} = i S_0$ through $\frac{\partial}{\partial \tau} = \nabla S_{0}^{E} \nabla$. This is consistent with the semiclassical definition of the Lorentzian action, $\frac{\partial}{\partial t}=\nabla S_{0} \nabla$ by identifying $\tau = i t$, at the end of the calculation when we analytically continue to Lorentzian time.The mass term entering the equation is defined with respect to the inverted potential $m^2 = -V'' = |V''| > 0$, and therefore is positive.

We now assume the following ansatz for the perturbations
\begin{equation}
\psi_{n}^{E} = N(\tau) e^{- \frac{\Omega^{E}_{n}}{2} f_{n}^{2}}
\label{euclidperturbation}
\end{equation}

Replacing this form in Eq. \ref{euclidfluctuationeq} leads to
\begin{equation}
\frac{\partial N}{\partial \tau} = - \frac{\Omega^{E}_{n}}{2 a^3}
\label{euclidN}
\end{equation}
 and
\begin{equation}
- {\dot \Omega^{E}_{n}} =\frac{ \Omega^{E 2}_{n} - \omega^{2}}{a^3}
\label{euclidomega}
\end{equation}

where the dot here, Eqn. \ref{euclidomega}, means $\frac{\partial}{\partial \tau}$. From the next line on we drop the indices $(n, E)$ unless it is neccessary to emphasize that we are performing the calculation in the Euclidean plane.
Considering
\begin{equation}
\Omega^{E} = \frac{a^{3} {\dot y}}{y}
\label{euclidomega1}
\end{equation}
 and replacing it in Eq. \ref{euclidomega} yields
\begin{equation}
\frac{d\Omega}{d\tau} = 3 a^{2} {\dot a} \frac{{\dot y}}{y} + a^{3}\left(\frac{{\ddot y}}{y} - \frac{{\dot y}^2}{y^2}\right) = \frac{\omega^{2} - \Omega^{2}}{a^3}
\end{equation}

This equation, in terms of a 'conformal' time defined by $d\tau = a d\eta^{E}$, becomes

\begin{equation}
y'' + 2 \frac{{a'}}{a} y' - \left(\frac{n^2}{a^2} + m^{2} \right) y = 0
\label{euclidmode}
\end{equation}

where prime is equal to $\frac{\partial}{\partial \eta^{E}} $. In the Lorentzian sector, a nearly exponential expansion with a Hubble constant ${\cal H}$ corresponds to a DS 3-geometry with scale factor $a(\eta) = -\frac{1}{{\cal H}\eta}$. In the Euclidean sector we can see that $\tau -> i t$ is equivalent to the transformation $a^{E}(\eta) = - i a(\eta) = \frac{i}{{\cal H} \eta}$ consistent with a Euclidean rotation, (or $\eta^{E} = i\eta)$. We now replace $a^{E}(\eta) = \frac{i}{{\cal H} \eta}$ in Eq. \ref{euclidmode} and drop the $^{E}$ notation,  to get
\begin{equation}
y'' -  \frac{2}{\eta} y' + \left(\frac{m^{2}}{{\cal H}^2 \eta^2 }- n^2 \right) y = 0
\label{euclidmode2}
\end{equation}

Eq. \ref{euclidmode2} can be solved exactly. First, let's write $ y = \eta^{3/2} Z(\eta)$, which leads to an equation for $Z(\eta)$

\begin{equation}
Z'' +  \frac{1}{\eta} Z' + \left[ \left(\frac{m^2}{{\cal H}^2} - \frac{9}{4}\right)\frac{1}{\eta^2} - n^{2}\right] Z = 0
\label{euclidmode3}
\end{equation}

Eq. \ref{euclidmode3} is solved by the modified Bessel functions $K_{\nu}(x)$ since the argument $x$ is imaginary, $ x = \kappa_{n} \eta $ with $\kappa_{n}^{2} = - n^2$, and $\nu^{2} = \frac{9}{4} - \frac{m^2}{{\cal H}^2} \simeq 9/4$. We can not choose the other modified Bessel function $I_{\nu}(x)$ that solves this equation since $I_\nu$  would give a purely real $\Omega$ which eventually leads to a decoherence factor $D=0$ when moving to the Lorentz sector. It should be noted that for $\nu = 1/2$ the choice $I_{\nu}$ corresponds to an exponentially growing function while $K_{\nu}$ to a decaying function.Therefore viewing Eqn.\ref{euclidfluctuationeq} as a Schrodinger equation of a wave in a potential well, the choice of the decaying solution $K_{\nu}$ is the physically relevant solution, since the other function, the exponentially growing solution denotes instability. Under the unitary evolution of Eqn. \ref{euclidfluctuationeq}, a decaying solution will evolve into a decaying solution and not an unstable exponentially growing one, therefore these solutions form a closed subset in the Hilbert space.

We need a solution that allows $\Omega$ to have a real and an imaginary part, since the reduced density matrix is obtained by the wavefucntion sqaured, traced over the fluctuation modes $f_n$. Therefore we 
obtain a decoherence factor (given by the real part in the exponent of the reduced density matrix) and an energy shift (the phase shift of the wavefunction from its classical path in midisuperspace, given by the imaginary part of the exponent of the reduced density matrix), only when $\Omega$ is complex.

Choosing $K_{\nu}(in\eta)$ allows $\Omega$ to be complex, therefore the suppression factor among the branches is not zero. $\Omega$ is imaginary for the other choice of the function $I_{\nu}(x)$.
We can use the transformation properties of $K_{\nu}$ in terms of the Hankel function of the first kind $H^{(1)}_{\nu}$

\begin{equation}
K_{\nu}(x) = \frac{1}{2} \pi i e^{\frac{i \nu \pi}{2}} H_{\nu}^{(1)} (x e^{\frac{i\pi}{2}})
\end{equation}
from which

\begin{equation}
K_{1/2}(x) = \frac{i\pi}{2}e^{\frac{i \pi}{4}} H_{1/2}^{(1)}(i x),  \qquad \nonumber 
K_{3/2}(x) = \frac{i\pi}{2} e^{\frac{i 3\pi}{4}} H_{3/2}^{(1)}(i x). 
\end{equation}
and thus with $x=i n \eta$ we have

\begin{equation}
 x = \frac{i n}{a {\cal H}} = \frac{n}{a^{E}{\cal H}},  \qquad \nonumber
 i x = - \frac{n}{a {\cal H}},  \qquad \nonumber
\frac{K_{1/2}}{K_{3/2}} = - i \frac{H^{(1)}_{1/2}}{H^{(1)}_{3/2}}.
\end{equation}
where
\begin{equation}
K_{\nu}(x)= K_{\nu}(\frac{i n}{a {\cal H}})= K_{\nu}(\frac{n}{a^{E}{\cal H}}),  \qquad \nonumber
H^{(1)}_{\nu}(i x) = H^{(1)}_{\nu}(\frac{-n}{a {\cal H}}).
\end{equation}
Finally from $\frac{K_{1/2}}{K_{3/2}} = -i \frac{H^{(1)}_{1/2} (-\frac{n}{a{\cal H}})}{H^{(1)}_{3/2}(-\frac{n}{a{\cal H}})}$ we get
\begin{equation}
\Omega^{E} = \frac{-1}{\eta^2 {\cal H}^2} \left[\frac{m^2}{3 \eta {\cal H}^2} + i n (-i)\frac{H^{(1)}_{1/2}(i x)}{H^{(1)}_{3/2}(i x)} \right]
\label{euclidomega}
\end{equation}

\subsection{Analytic Continuation}

Applying the WKB approximation near the saddle point of the inverted potential in the Euclidean sector, $a^{E}(\eta)= -i a(\eta)$ allows us to write the total wavefunction as: $\Psi^{E} \simeq e^{-\frac{S_{0}}{\hbar}} \psi_{n}^{E}$ with $\psi_{n}^{E}= N(\tau)e^{-\frac{\Omega^{E}}{2} f_{n}^{2}}$. From Eqs. \ref{omegaeq} and Eq. \ref{euclidomega1} we can see that 
\begin{equation}
\Omega^{E} = - i \Omega = -i \left(\Omega_{R} +i \Omega_I{} \right) = \Omega^{E}_{R} + i \Omega^{E}_{I} 
\label{omegarelate}
\end{equation} 
Therefore when rotating back to the Lorentz time $t$, the real part of $\Omega^{E}$ which in Euclidean space contributes to the energy shift, becomes the imaginary part of its Lorentzian equivalent and vice versa from Eq. \ref{omegarelate}. From Eq. \ref{omegarelate} and using the solution of Eq. \ref{euclidomega}, we have the following identification: $Re(\Omega^{E}) = \Omega^{E}_{R}= +i Im(\Omega) = \left(\frac{n^{2}a^{3} {\cal H}}{n^{2} + a^{2} {\cal H}^{2}} + \frac{m^{2} a^{3}}{3 {\cal H}} \right)$ and $Im(\Omega^{E})= \Omega^{E}_{I} = - Re(\Omega) = - \frac{n^{3} a^{2}}{n^{2}+ a^{2}{\cal H}^2}$. (Note that the sign of the energy correction term given by $Im(\Omega^{E})$ is inverted to $ - Re(\Omega)$ when going back to the Lorentzian sector). 

 We thus have the following result for the total effective potential and the energy shift induced by fluctuation that are driven by $m^2 = |V''| >0$, when moving back from Euclidean to the Lorentz sector, with respect to real time $t$, and the $dot$ now being given by the 'Lorentzian' $d/dt = i d/d\tau$ 
 
\begin{equation}
V^{E}_{eff}(\phi)= - V(\phi) - \delta E^{E}_{\phi}(\phi, b) = - V_{eff}
\end{equation}

where now,

\begin{equation}
 V_{eff} = V(\phi) + \delta E_{\phi}(\phi, b)
\label{lorentzv}
\end{equation}

since $V^{E} \rightarrow - V(\phi)$ and,

\begin{equation}
\delta E_{\phi}^{E} \rightarrow - \delta E_{\phi} = + \frac{1}{a^3}\left[Tr(Re[\Omega]) + Tr(\frac{i d\Omega_{I}/d(it)}{2\Omega_{R}})\right]
\label{concaveenergy}
\end{equation}

So we have just shown that applying the WKB approximation for concave potentials with $V''<0$ required some subtlety involving a rotation to the Euclidean sector. Back in the Lorentz plane, not only is the effective potential energy sign flipped, for both $V(\phi$ and $\delta E_{\phi}$, but the energy shift for concave potentials is induced by fluctuations that are driven by $|V''|$ in Eq.\ref{lorentzv}, i.e from fluctuations around a saddle point of the inverted 'convex' potential with $m^2 > 0$.
It can be seen that the decoherence factor, $Exp[-D]$, back in the Lorentzian sector, with the identifications above of $Re[\Omega]=\Omega_{R}= -\Omega^{E}_{I}$ becomes
\begin{equation}
2D = Tr(\frac{\Omega_I}{\Omega_{R}})^{2} = \Sigma_{n} n^{2} \left(\frac{\Omega_{I}}{\Omega_{R}} \right)^{2}
\label{decoherence}
\end{equation}

i.e. it remains exactly the same for both, concave and convex potentials. The integrals and summations of these expressions are calculated in the Appendix (see also \cite{richlmh} for convex potentials).

 After the emergence of a classical universe, the energy shift in the concave potential Eq. \ref{concaveenergy}, contributes to the Friedmann equation by modifying the inflationary concave downward potential as follows
\begin{equation}
V^{concave}_{eff}(\phi)= V^{concave}(\phi) + \delta E_{\phi}
\label{concaveenergy}
\end{equation}
which is to be contrasted to the energy correction for convex potentials calculated directly in the Lorentz sector illustrated with an exponential potential in \cite{richlmh}.
 
 Defining the entanglement length $L_{i}(k, b)$ as in \cite{tomolmh}, we can now obtain the modification to the Newtonian potential of the universe $\Phi_{0}$, from the nonlocal modification $\delta E(\phi, b)$ to the effective potential, by using the Poisson equation $\nabla^{2} \delta \Phi = 4\pi G_{N} \delta E(\phi, b) $, where $\Phi_{0}$ is obtained from $V$ and $\Phi$ from $V_{eff}$ since $-\frac{2}{3} (\frac{k}{aH})^{2} \Phi = 4 \pi G_{N}$. Thus, with $L_i$ defined in  \cite{tomolmh} we have 
 
 \begin{equation}
 \Phi = \Phi_0 + \delta \Phi \approx \Phi_{0} \left[ 1 + \frac{\delta E(\phi, b)}{\rho} ( \frac{r}{L_{i}(k, b)} )^{2} \right]
 \label{newtonpotential}
 \end{equation}
 
 Note that the correction to the effective potential $\delta E$ is negative.
 The giant void and the other anomalies arise from the correction $\delta \Phi$ of the Newtonian potential, discussed in \cite{tomolmh}. We present in detail the implication of this modification to $\Phi$ and the comparison of the anomalies nested in this effective potential, modified Newtonian potential, and the modified field solutions, against the best fit parameters of the data from the Planck satellite soon in \cite{eleonora}.
 
Although the real part of the exponent in the wavefunction (or equivalently in the reduced density matrix) in the Lorentz frame gives the same decoherence factor $e^{-D}$ for concave potentials as for the convex case, it is important to remember that the curvature of the inverted potential driving the fluctuation and determining the mass term in the equations above is given by $m^2 =|V''|>0$, not by $m^2 = V''<0$, and it remains so after rotating from Euclidean to Lorentz sectors.

The method presented here of an Euclidean rotation for the case of fluctuation around concave potentials with $V'' <0$, is a similar technique to the one used for instantons in quantum mechanics. The results presented here for the energy correction, decoherence and the definition of $m^2>0$ are applicable to all concave downward potentials with $V''<0$. The main difference with the case of convex potentials which have stable fluctuations since $V''>0$ is the sign of the energy correction calculated from entanglement in the landscape: the energy shift is added to the energy of concave potentials and it is subtracted from that of convex potentials. Decoherence is the same in both cases with the understanding that $m^2 =|V''|>0$ always.

We will apply the signatures derived here which originate from the quantum entanglement of our branch of the wavefunction with others in the quantum multiverse, to two examples of concave downward potentials: the Starobinsky model and the hilltop model, as well as revisit the convex potential of \cite{tomolmh} in the light of the new data from Planck satellite in two companion papers \cite{eleonora}.

\section{Conclusions}
\label{sec:conclusions}

We have extended the applicability of our theory of the origin of the universe from a quantum multiverse to a class of inflationary models which are subtle and require careful treatment, because a straightforward application of semiclassical methods is not possible.

We derived the effect of quantum entanglement of our branch universe with other branches of the wavefunction of the universe, for the case of concave downwards inflationary potentials with $V''<0$ when the field is in a part of its potential away from an extrema, thus the semiclassical approach fails. The reason is: expansion of the action around the saddle point can not be justified since the slow rolling field is in a part of the potential away from the saddle point and fluctuations are driven by $V''<0$. 

Here we performed the flucutations calculation and the WKB expansion in the Euclidean sector which inverts the potential. In the inverted potential, fluctuations are driven by $m^2 = -V'' >0$, and a semiclassicalexpansion around the potential 'well' of the inverted potential is justified.

Inverting the potential, by moving to the Euclidean sector,
 is justified from an inherent symmetry of quantum cosmology \cite{vilenkin, hawking}. At the end we rotate back to Lorentzian geometries and find that the energy shift is added to the concave potential, in contrast to being substracted from convex potentials, but the decoherence factor remains the same. We also show why naively considering $m^2 = V'' <0$ and from there perform a WKB expansion to
estimate the effect of fluctuations in Lorentzian sector, is erronous and produces a growing instead of suppressing decoherence factor.

We will report in a companion publications, how the list of modifications to the Newtonian potential and the Friedmann equation derived here from entanglement comapres with the new data from Planck satellite.

\section{Appendix}

Here we show the summation involved in all the trace terms that enter the expression for the energy shift, interference length, and for the decoherence factor, performed after rotation in the Lorentz plane.

In what follows we replace the infinite sum with an integral over all the modes $b<\frac{n}{a}<{\cal H}$ with comoving wavelengths from the horizon size $a{\cal H}$ to the wavelengths which probe the coherence of the wavepacket in the midi-superspace. Those are the modes whose inverse wavelength corresponds the width of the wavepacket (or branch), $a b$ \cite{tomolmh}.

The interference length is the length at which the quantum nature of the universe such as the entanglement with other branches becomes important, and can be read off from the imaginary term in the reduced density matrix $e^{-i Tr(\frac{\Omega_I}{\Omega_R}) - D}$. It is given by $A_{1}/{\cal H}$, where 

\begin{equation}
A_{1} = Tr(\frac{\Omega_I}{\Omega_R}) \approx \int_{ab}^{a{\cal H}} dn a\left[\frac{m^{2}(n^{2} + a^{2}{\cal H}^2) + 3 n^{2}{\cal H}^{2}}{3n^{3}{\cal H}}\right] \simeq a \left[-(\frac{m^{2}}{3{\cal H}} + 3{\cal H})Log(b/{\cal H}) + \frac{m^{2}{\cal H}}{6 b^2}Exp[-\frac{b^2}{{\cal H}^2}]\right]
\label{a1}
\end{equation}
The real part in the exponent of the density matrix gives the decoherence factor as in Eq.\ref{decoherence}

\begin{equation}
D = \frac{1}{2} \Sigma_{n} n^2 (\frac{\Omega_{I}}{\Omega_R})^2 \approx \int_{ab}^{a{\cal H}} n dn a^2 \left[-(\frac{m^{2}}{3{\cal H}} + 3{\cal H})Log(b/{\cal H}) + \frac{m^{2}{\cal H}}{6 b^2}Exp[-\frac{b^2}{{\cal H}^2}]\right]^2 
\end{equation}

which is a simple integral but long. So if we focus on the limit where $a$ grows, $ma>1$ for example, then the above is roughly: $e^{-D}\simeq e^{-m^{3} \frac{a {\cal H}^2}{b^{2}}(a_{1} -a_{2})^2} \simeq e^{-10^50}$ very efficient decoherence.

Now we go on to estimate
\begin{equation}
A_{3} =  Tr(\Omega_R) = \int_{ab}^{a{\cal H}}  n dn \frac{n^{2}a^{2}}{n^{2} + a^{2}{\cal H}^2} = \frac{a^{4}{\cal H}^2}{2}\left[(\frac{b^2}{{\cal H}^2} - 1) - Log[b/{\cal H}] \right]
\label{a3}
\end{equation}
Finally we estimate the last term that contributes to the energyshift below, by using ${\dot \Omega_I} = (a{\cal H})( d\Omega_{I} / da)$, so,
\begin{equation}
A_{4} = Tr(\frac{{\dot \Omega_{I}}}{2 \Omega_{R}} )=\frac{3{\cal H}}{2}A_{1} - \int dn \frac{(a{\cal H})^4}{na}(\frac{1}{(n^2 + a^2{\cal H}^{2}})
\end{equation}

We now have the energy shift for concave potentials, which modulo the sign in front of it and the definition of $m^{2}=Abs[V'']$ ,  is the same as the correction calculated for convex potentials in \cite{tomolmh}. Using the Friedmann equation to relate $V_{eff}(\phi)$ to ${\cal H}^2$ allows us to write
\begin{equation}
\delta E_{\phi} = \frac{1}{a^3}\left[A_{3} + A_{4}\right] = + \frac{V(\phi)^2}{18 M^4} F[b,V(\phi)] \\
\end{equation}

where 

\begin{equation}
F[b,V] =\left[\frac{3}{2} (2+ \frac{m^2}{3{\cal H}^2})Log[\frac{M^2 b^{2}}{V}] - \frac{1}{2}(1 + \frac{m^2}{b^2})e^{-3 \frac{b^2 M^2}{V}} \right]
 \end{equation}

In summary, with $m^2 = |V''|$ for concave potentials we find 
\begin{equation}
3 {\cal H}^2 M^2 = V_{eff}^{concave}= V^{concave} +  \frac{V^{concave}(\phi)^2}{18 M^4} F[b,V^{concave}(\phi)] \nonumber
\end{equation}

and for convex potentials with $m^2 = V''$ we had \cite{tomolmh} 

\begin{equation}
3 {\cal H}^2 M^2 = V_{eff}^{convex}= V^{convex} - \frac{V^{convex}(\phi)^2}{18 M^4} F[b,V^{convex}(\phi)] = V^{convex} +  |\frac{V^{convex}(\phi)^2}{18 M^4} F[b,V^{convex}(\phi)]|.
\label{f}
\end{equation}

Note that $F[b,V]$ is always negative, thus $ - F[b,V] = |F[b,V]|$ used in the last line of Eq.\ref{f}. \\

\acknowledgments
LMH acknowledges support from the Bahnson funds.





\end{document}